# Interplay of Dirac electrons and magnetism in AMnBi$_2$ (A=Ca, Sr)

Anmin Zhang,[1] Changle Liu,[1] Changjiang Yi,[2] Guihua Zhao,[1] Tian-long Xia,[1] Jianting Ji,[1] Youguo Shi,[2] Rong Yu,[1,5] Xiaoqun Wang,[1,4,5] Changfeng Chen,[3] and Qingming Zhang[1,5]*

[1]Department of Physics, Beijing Key Laboratory of Opto-Electronic Functional Materials & Micro-nano Devices, Renmin University of China, Beijing 100872, P. R. China
[2]Beijing National Laboratory for Condensed Matter Physics, Institute of Physics, Chinese Academy of Sciences, Beijing 100190, P. R. China
[3]Department of Physics and High Pressure Science and Engineering Center, University of Nevada, Las Vegas, Nevada 89154, USA
[4]Department of Physics and Astronomy, Shanghai Jiao Tong University, Shanghai 200240, P. R. China
[5]Collaborative Innovation Center of Advanced Microstructures, Nanjing 210093, P. R. China
*Corresponding author: qmzhang@ruc.edu.cn

## Abstract

Dirac materials exhibit intriguing low-energy carrier dynamics that offer a fertile ground for novel physics discovery. Of particular interest is the interplay of Dirac carriers with other quantum phenomena, such as magnetism. Here we report on a two-magnon Raman scattering study of AMnBi$_2$ (A=Ca, Sr), a prototypical magnetic Dirac system comprising alternating Dirac-carrier and magnetic layers. We present the first accurate determination of the exchange energies in these compounds and, by comparison to the reference compound BaMn$_2$Bi$_2$, we show that the Dirac-carrier layers in AMnBi$_2$ significantly enhance the exchange coupling between the magnetic layers, which in turn drives a charge-gap opening along the Dirac locus. Our findings break new grounds in unveiling the fundamental physics of magnetic Dirac materials, which offer a novel platform for probing a distinct type of spin-Fermion interaction. The outstanding properties of these materials allow a delicate manipulation of the interaction between the Dirac carriers and magnetic moments, thus holding great promise for applications in magnetic Dirac devices.

**Introduction**

Recent years have seen the emergence of a new class of materials whose low-energy carrier dynamics obey the Dirac equation, instead of the Schrodinger equation that describes most condensed matter systems. These so-called Dirac materials exhibit linear carrier dispersion and massless chiral excitations that give rise to novel quantum phenomena such as the ultra-high electron mobility and quantum Hall effect[1-4]. So-far identified Dirac materials include graphene[2,3], topological insulators[5,6], and d-wave and iron-pnictide superconductors[7,8].

One of the most intriguing aspects of Dirac materials is the interplay of their unique carrier dynamics with other quantum phenomena. A prominent case is magnetism, which may significantly change the electronic band structures of Dirac materials, as demonstrated by an antiferromagnetic (AF) long-range order in graphene observed in a recent experiment[9]. Further studies on the mutual influence of the Dirac-type electronic excitations and magnetism are impeded by the small size of available graphene samples (~7 nm in width). A suitable model material system that possesses both Dirac carriers and magnetic order is essential to further exploration of novel physics and innovative device concepts in magnetic Dirac materials. The recent discovery of coexisting linear Dirac bands and long-range magnetic order in $SrMnBi_2$ and $CaMnBi_2$ provides an exciting platform for the study of magnetic Dirac materials.

Transport measurements reveal linear band dispersions near the Fermi energy in both $AMnBi_2$ (A=Sr, Ca) compounds[10-13]. First-principles calculations indicate that such Dirac-type linear dispersions come from the $6p_x$ and $6p_y$ orbits of the Bi ions in the intercalated Ca(Sr)Bi layers, which are slightly hybridized with the $d$ orbits of the Ca or Sr ions[10,12,14,15]. The calculated band structure has been verified by angle-resolved photoemission (ARPES) experiments[10,16,17]. The transport measurements also indicate an AF transition around 290 K[10-13]. The ground state predicted by first-principles calculations has a checkerboard AF order of $Mn^{2+}$ spins, with a spin moment of ~4 $\mu_B$[14,15]. Subsequent neutron diffraction measurements[18] confirmed the AF transition and the estimated size of the magnetic moment, and the experiment also demonstrated that CaMnBi$_2$ and SrMnBi$_2$ have the C-type and G-type AF structures, respectively. Moreover, it is shown that magnetic ordering could open an energy gap in Dirac fermion band in CaMnBi$_2$ but not in SrMnBi$_2$ at the mean-field level[18]. The AMnBi$_2$ compounds comprise alternating Ca(Sr)Bi-layers accommodating 2D Dirac electrons and MnBi-layers containing a long-range magnetic order, and this configuration is similar to the case reported in the magnetic graphene. The availability of large-size AMnBi$_2$ crystals allows experimental explorations of the interplay between the coexisting magnetic order and Dirac electrons.

In this work, we report on Raman scattering measurements of two-magnon excitations in SrMnBi$_2$ and CaMnBi$_2$. From the measured and calculated Raman spectra, we have determined, for the first time, the nearest-neighbor ($J_1$), next-nearest-neighbor ($J_2$) and interlayer ($J_c$) exchange energy. By comparison to a reference system BaMn$_2$Bi$_2$, we find an unusual enhancement of the interlayer exchange couplings between neighboring magnetic layers via the intervening Dirac-carrier layers. We further examined the effect of the magnetism on the Dirac electron band structure using a spin-fermion model. Our results show that the enhanced interlayer exchange coupling drives a charge-gap opening along the Dirac locus in CaMnBi$_2$. Unlike the effect of the spin-orbit coupling (SOC), the gap opened by the AF ordering allows both the upper and lower branches of the Dirac locus to cross the Fermi level, which well explains recent ARPES measurements[16]. The present study addresses a fundamental issue in magnetic Dirac materials, i.e., the interplay between Dirac carriers and magnetism. The materials systems studied here provide a novel platform for probing a new type of charge-moment interaction where the magnetic and conducting layers are coupled but well separated, making these materials a new family of prototypes well described by the spin-Fermion model. This unusual separation of conducting (Dirac) charge and magnetic moments allows a delicate manipulation of the interaction between the charge and moment subsystems, which can be explored for innovative spintronic applications.

## Results

Identification of two-magnon Raman spectra

The three compounds CaMnBi$_2$, SrMnBi$_2$ and BaMn$_2$Bi$_2$ share similar crystal and magnetic structures, as shown in Fig. 1a-1c, and their Raman spectra exhibit a common feature around 500~800 cm$^{-1}$ (Fig. 1d). The measured spectral weights shift towards larger wavenumbers following the order of the ionic radii of Ca, Sr and Ba, and the overall spectral feature remains unchanged under different excitation sources (see the inset in Fig. 1d), which indicates that these spectra come from the Raman process rather than a photoluminescence process. A multi-phonon process is also unlikely because there are no strong phonon excitations above 300 cm$^{-1}$. Moreover, we analyzed the two-magnon Raman process in BaMn$_2$Bi$_2$, using the exchange couplings $SJ_1$=21.7(1.5) meV, $SJ_2$=7.85(1.4) meV, $SJ_c$=1.26(0.02) meV determined by the neutron scattering measurements[19]. Our calculations (see Supplementary Note 1 for the method of calculation) produced two-magnon excitations that peak around 650 cm$^{-1}$, in good agreement with our experimental observation. This agreement between experiment and theory further demonstrates that the spectral feature around 500-800 cm$^{-1}$ in these Mn-Bi compounds originates from the two-magnon Raman process.

Evolution of two-magnon peak with temperature

We show in Fig. 2 the evolution of the two-magnon peak with temperature. The spectra from all three compounds exhibit the same trends with increasing temperature, including a shift to lower wavenumbers in peak position, a gradual broadening in peak width, and a reduction in intensity, which nevertheless remains visible above the transition temperatures ($T_N$). These trends are typical for the two-magnon Raman process. The shift of the peak position reflects the energy changes of the large-q magnons and magnon-magnon interactions, the peak broadening indicates a decrease of the magnon lifetime, and the visibility of the peak structure even above $T_N$ follows a general feature of the two-magnon process, since the peak is dominated by the magnons at the Brillouin zone boundary where the magnetic correlation remains viable even far above $T_N$[20]. The peak at ~500 cm$^{-1}$ is likely associated with a process involving a phonon and a magnon, considering its temperature evolution is similar to that of the two-magnon spectra and there is a strong spin-orbit coupling in the system[21]. The anomalies in resistivity and susceptibility reported at 260 K and 50 K in SrMnBi$_2$ and CaMnBi$_2$, respectively, were attributed to spin realignments or a slight structural change without any anomalies in specific heat[12-14,18]. We observed no anomalies in the two-magnon spectra at these temperatures (Figs. 2a&b). Our results thus rule out spin-related processes as the origin of these anomalies since two-magnon spectra are highly sensitive to variations of the magnetic order in the system.

Determination of exchange energies

By comparing the characteristic energies extracted from the spectra to the calculated two-magnon density of states (DOS), we have determined, for the first time, the exchange energies in SrMnBi$_2$ and CaMnBi$_2$, as summarized in Table I (See Supplementary Notes 2-7 , Supplementary Figs. 1-3 and Supplementary Tables 1-3 for details). We also have extracted the exchange energies for BaMn$_2$Bi$_2$, and the values are in good agreement with those determined by the neutron scattering measurements[19], which confirms the reliability of the results from the two-magnon Raman spectra. The small difference between the $J_c$ values extracted by the two methods may have resulted from different sample sources and/or experimental methods.

The very high energy resolution of Raman scattering (~ 0.1 meV) combined with the sharp two-magnon features observed in all three crystals are expected to produce an accurate determination of the magnetic exchange energy[25]. There is also appropriate verification on the validity of the theoretical model adopted in the present work, which predicts that the characteristic points in two symmetry channels are exactly the same (Fig. 3, $A_{1g}$ and $B_{1g}$, see below). This predicted coincidence is indeed seen in the experimental spectra. There are four available characteristic points in the two-magnon spectra, and any three of them can provide the same exchange parameters. We have examined the accuracy of the extracted exchange energies by inputting much larger error bars for the raw data (Supplementary Notes 6 and 7, Supplementary Figs 2 and 3). Particularly for the small $J_c$, its value is exactly proportional to the width of the plateau between $\omega_1$ and $\omega_2$ (Supplementary Note 2). The value of $J_c$ read out from Fig. 3 well coincides with those listed in Table I. It should be emphasized that the same modeling is equally applied to all three systems studied in the present work without any additional constraints. This means that the exchange energies relative to each other are consistently comparable even if their magnitudes may have deviations.

Based on the obtained exchange energies, we have calculated the two-magnon spectra in the $A_{1g}$ and $B_{1g}$ channels (see Supplementary Note 1), where the irreducible representations of the $D_{4h}$ point group, $A_{1g}$ and $B_{1g}$, denote different symmetry channels and can be separated by configuring the polarizations of incident and scattered light. The results (Fig. 3) show that the magnon-magnon interactions have little influence on the $A_{1g}$ spectra but contribute a sharp resonant peak in the $B_{1g}$ channel. Magnon-magnon scattering drives a spectral weight transfer to lower energies and consequently causes such a magnetic exciton-like resonance peak, whose position corresponds to the exciton energy. The simulations including the magnon-magnon interactions produce results in much better agreement with the experimental data for $BaMn_2Bi_2$ compared to the results of the non-interacting calculations. However, the non-interacting results agree better with the experimental spectra in $AMnBi_2$. It should be noted that the non-interacting calculations in $AMnBi_2$ have some discrepancies with the experimental data in some spectral details. The presence of itinerant electrons, particularly in $CaMnBi_2$ and $SrMnBi_2$, may be responsible for such deviations. The itinerant electrons tend to reduce the effective intensities of the incident light and contribute to a relatively low signal-to-noise ratio. Furthermore, itinerant electrons also bring higher-order corrections to the linear spin-wave model. Theoretical calculations for such corrections are extremely complicated and have not been archived in the literature; however, such corrections are not expected to affect the main features and characteristic points in the spectra, although they can possibly modify some spectral details.

Discussions

The strong suppression of the resonant peak in Ca(Sr)MnBi$_2$ is associated with the strong SOC effect in the Ca(Sr)Bi layer,[16] which leads to an easy-axis anisotropy (along the $S^z$ direction) of the exchange couplings, as well as an anisotropic damping of the magnon-magnon interactions that can suppress the resonant peak in the B$_{1g}$ channel of the Raman spectra. Although the resonance stems from magnon-magnon interactions as mentioned above and is strongly suppressed by the SOC effect, the resonance peak intensities may not be a good measure of the interaction or SOC strengths. This is because many basic factors such as distinguished crystal symmetries (the presence of a horizontal mirror plane in SrMnBi$_2$ but absent in CaMnBi$_2$), magnetic structures (G-type for SrMnBi$_2$ but C-type for CaMnBi$_2$) and carriers (Dirac electrons dominant in SrMnBi$_2$ but Dirac plus ordinary electrons in CaMnBi$_2$) are not taken into account. Further insights into this important issue require additional experimental and theoretical investigation.

The in-plane exchange couplings $J_1$ and $J_2$ in the three compounds studied here are well correlated with the in-plane lattice constants. SrMnBi$_2$ has the largest lattice parameter along its *a* axis and the smallest $J_1$ and $J_2$; BaMn$_2$Bi$_2$ and CaMnBi$_2$ have similar *a* values and their $J_1$ and $J_2$ are quite close to each other (see Table I). This close correlation suggests that the in-plane magnetism is well described by the super-exchange mechanism.

Table I   The magnetic exchange energies extracted from the Raman spectra. Here *a* is the in-plane lattice constant and *d* the distance between the neighboring MnBi-layers.

|  | $SJ_1$ (meV) | $SJ_2$ (meV) | $SJ_c$ (meV) | $a$ (Å) | $d$ (Å) |
|---|---|---|---|---|---|
| **CaMnBi$_2$** | 20.77(0.79) | 7.29(0.48) | -1.31(0.10) | 4.50[a] | 11.07[a] |
| **SrMnBi$_2$** | 16.00(0.30) | 4.75(0.17) | 2.92(0.09) | 4.58[a] | 11.57[a] |
| **BaMn$_2$Bi$_2$** | 21.45(0.32) / 21.7(1.5)[b] | 6.26(0.20) / 7.85(1.4)[b] | 0.78(0.08) / 1.26(0.02)[b] | 4.49[b] | 7.34[b] |

a: ref. 18; b: ref. 19

In sharp contrast, the interlayer coupling $J_c$ exhibits highly anomalous behavior. The distances between the neighboring MnBi-layers in CaMnBi$_2$ and SrMnBi$_2$ (11.07 Å and 11.57 Å, respectively) are much larger than that in BaMn$_2$Bi$_2$ (7.34 Å) because of the intercalation of the additional Bi-layers in the two magnetic Dirac compounds. At such large interlayer distances, the interlayer coupling $J_c$ is usually expected to be negligible as suggested by recent neutron measurements and calculations[15,18]. In fact, negligible $J_c$ values have been reported in many other layered compounds with magnetic interlayer distance ≳0.7 nm, such as K$_2$NiF$_4$, K$_2$MnF$_4$ and Rb$_2$MnF$_4$[23,24]. Surprisingly, the extracted $J_c$ value for SrMnBi$_2$, which has the largest MnBi-interlayer distance, is 3.6 times that in BaMn$_2$Bi$_2$, which has the smallest MnBi-interlayer distance. Meanwhile, CaMnBi$_2$ also has a larger $J_c$ about 1.7 times that of BaMn$_2$Bi$_2$. This unusual enhancement of $J_c$ is apparently beyond the standard super-exchange mechanism and indicates novel physics in these magnetic Dirac materials. A key structural feature in the AMnBi$_2$ compounds is a Dirac-carrier Ca(Sr)Bi layer between the neighboring magnetic MnBi-layers; in contrast, there is only a single layer of Ba$^{2+}$ ions between the neighboring MnBi-layers in BaMn$_2$Bi$_2$. This structural contrast suggests that the enhanced interlayer magnetic coupling stems from the interplay of magnetism and the Dirac carriers in the intervening Ca(Sr)Bi layer. The spin-fermion systems studied here provide unique insights into the novel physics of the composite magnetic and Dirac electron systems where the layers accommodating itinerant carriers are sandwiched by ordered and insulating magnetic layers. On the other hand, the present material systems do not provide an adequate platform to clearly identify the role of the ordinary electrons. A definitive resolution of this issue requires the synthesis of appropriately structured spin-fermion systems and additional theoretical exploration, which is beyond the scope of our present work.

To understand the interplay of the Dirac carriers and magnetism in SrMnBi$_2$ and CaMnBi$_2$, especially the enhancement of the coupling $J_c$ between neighboring magnetic layers and the modifications of the electronic band structure in the Dirac-carrier layers, we consider the following effective spin-fermion model describing both the itinerant electrons in the Bi $6p_x$ and $6p_y$ orbits and interacting local magnetic moments on the Mn ions:

$$H = \sum_{i,j,\alpha,\beta,l} t_{ij}^{\alpha\beta l} c_{i\alpha l}^+ c_{j\beta l} + \lambda_{SO} \sum_{i\alpha\beta ll'} c_{i\alpha l}^+ c_{i\beta l'} + \frac{J_K}{2} \sum_{i\alpha ll'} c_{i\alpha l}^+ \sigma_{ll'} c_{j\alpha l'} \cdot S_{i\pm\hat{z}} + \sum_{i'j'} J_{i'j'}^H S_{i'} \cdot S_{j'} \quad (1)$$

where $c_{i\alpha l}^+$ creates an itinerant electron at site i in orbit α with spin index l in the Ca(Sr)Bi Dirac-carrier layer, $S_{i'}$ refers to the local moment of the Mn ion below or above the Ca(Sr)Bi layer, $t_{ij}^{\alpha\beta l}$ is the hopping integral of the itinerant electrons, $\lambda_{SO}$ is the spin-orbit coupling, $J_K$ is the Kondo coupling between the itinerant electrons and local moments, and $J^H$ is the super-exchange coupling between the local moments. Here we use the exchange energies determined from our two-magnon Raman spectra and treat the local moments of the Mn ions as classical spins, which is justified by the large moment of ~4 μ$_B$ per Mn at low temperatures, and the magnetic interaction in the AFM state is treated in a mean-field approximation.

The results of our model calculations show that both SrMnBi$_2$ and CaMnBi$_2$ possess anisotropic Dirac bands, but the processes for the gap opening between the upper and lower Dirac bands are very different in these two compounds. As already noticed in a previous study[15], the different arrangement of the Ca or Sr cations leads to a gap opening along a general direction in SrMnBi$_2$, but not in CaMnBi$_2$. As a result, there are four isolated anisotropic Dirac points along the Γ-M direction in SrMnBi$_2$, but a line of continuous Dirac points is present in CaMnBi$_2$ (see Supplementary Note 8, Supplementary Fig. 4 and Supplementary Table 4). We examine the effects of SOC and magnetic order on the gap opening in the Dirac bands. For SrMnBi$_2$, as shown in Fig. 4(a)-(b), the SOC opens a small gap (~0.01 eV for $\lambda_{SO}$=0.6 eV) at the Dirac band along the Γ-M direction, and it slightly enhances the existing gap between the lower and upper Dirac bands. The magnetic order has no net effect on the band structure at the mean-field level since the influence coming from the upper and lower Mn layers exactly cancel out due to the G-AFM order. For CaMnBi$_2$, the effect of SOC is similar, which opens a small gap of ~0.01 eV between the upper and lower Dirac bands. This gap, however, is much smaller than observed in a recent ARPES measurement, which is about 0.05-0.1 eV.

Surprisingly, we find that in CaMnBi$_2$, the C-AFM order introduces a massive term proportional to the sublattice magnetization, $J_K|m^z|$, which acts on itinerant electrons, and this term is highly effective in opening a gap in the Dirac bands (Fig. 4(c)-(d)). Taking $J_K$=0.01 eV, the gap increases five-fold to 0.05 eV, which is consistent with the value observed in recent ARPES experiment[16]. Meanwhile, we also estimated the effective $J_c$ in CaMnBi$_2$ driven by the RKKY interaction (see Supplementary Note 9). At the same $J_K$=0.01 eV, we obtained $|J_c|$~1 meV, which is in good agreement with the value obtained independently from fitting the Raman spectra (Table I). These results show consistently that the coupling between the Dirac electrons and local moments has a profound impact on both the effective interlayer magnetic interaction and the Dirac electronic band structure. This finding highlights a powerful characteristic of these magnetic Dirac systems and raises exciting prospects of manipulating these key properties by tuning the interlayer exchange coupling in magnetic Dirac materials. The spin-Fermion model traditionally applies to systems with magnetism and conducting carriers coexisting in the same lattice. The MnBi compounds studied here present a new environment where the magnetic moments and conducting carriers are well separated in different subsystems. This allows an accurate description of the spin-Fermion interaction in these materials, and the results offer new insights into the fundamental physics that may inspire innovative design concepts for spintronic applications.

In summary, we have performed a systematic two-magnon Raman study of magnetic Dirac compounds SrMnBi2 and CaMnBi2. Our measurements combined with model calculations produced, for the first time, an accurate determination of the exchange energies, which allow a quantitative understanding of the novel physics in these materials. A comparison with the reference compound $BaMn_2Bi_2$ reveals that the interlayer exchange couplings are significantly enhanced and the magnon-magnon interactions are suppressed by the Dirac-carrier layers. We further investigated the effects of magnetism on the band structure of Dirac carriers and found that the magnetic order has drastic effects on the gap opening in the Dirac bands in $CaMnBi_2$, which explains recent ARPES measurements. The discovery of the intriguing interplay of Dirac carriers and magnetism sheds new light on the rich physics in magnetic Dirac materials. Our reported work sets key benchmarks for these distinct systems containing coupled but well separated magnetic and Dirac-carrier layers that can be accurately described by the spin-Fermion model. These results unveil new fundamental physics and pave the way for innovative design and development of magnetic Dirac devices for spintronic applications.

**Methods**

High-quality crystals of $BaMn_2Bi_2$, $SrMnBi_2$ and $CaMnBi_2$ were grown by self-flux method. The details of crystal growth can be found elsewhere.[11,14,18,22] The antiferromagnetic transition temperatures of $SrMnBi_2$ and $CaMnBi_2$ could be find in Ref. 18, where magnetic susceptibilities and resistivities were measured in the same batch of crystals as used in our measurements. Raman measurements were performed with a Jobin Yvon HR800 single-grating-based micro-Raman system equipped with a volume Bragg grating low-wavenumber suite, a liquid-nitrogen cooled back-illuminated CCD detector and a 633 nm laser (Melles Griot). The laser was focused into a spot of ~5 μm in diameter on sample surface, with a power less than 100 μW to avoid overheating.

**References**

1. Vafek, O. & Vishwanath, A. Dirac fermions in solids-from high Tc cuprates and graphene to topological insulators and Weyl semimetals. *Annu. Rev. Condens. Matter Phys.* **5,** 83-112 (2014).
2. Novoselov, K. S. *et al.* Electric field effect in atomically thin carbon films. *Science* **306,** 666–669 (2004).
3. Castro Neto, A. H., Guinea, F., Peres, N. M., Novoselov, K. S. & Geim, A. K. The electronic properties of graphene. *Rev. Mod. Phys.* **81,** 109–162 (2009).
4. Zhang, Y., Tan, Y. W., Stormer, H. L. & Kim, P. Experimental observation of the quantum Hall effect and Berry's phase in graphene. *Nature* **438,** 201–204 (2005).
5. Hasan, M. & Kane, C. Colloquium: Topological insulators. *Rev. Mod. Phys.* **82,** 3045–3066 (2010).
6. Qi, X. & Zhang, S. Topological insulators and superconductors. *Rev. Mod. Phys.* **83**, 1057–1110 (2011).
7. Orenstein, J. & Millis, A. J. Advances in the physics of high-temperature superconductivity. *Science* **288,** 468–474 (2000).
8. Richard, P. *et al.* Observation of Dirac cone electronic dispersion in $BaFe_2As_2$. *Phys. Rev. Lett.* **104,** 137001 (2010).
9. Magda, G. Z. *et al.* Room-temperature magnetic order on zigzag edges of narrow graphene nanoribbons. *Nature* **514**, 608-611 (2014).
10. Park, J. *et al.* Anisotropic Dirac fermions in a Bi square net of $SrMnBi_2$. *Phys. Rev. Lett.* **107,** 126402 (2011).
11. Wang, K., Graf, D., Lei, H., Tozer, S. W. & Petrovic C. Quantum transport of two-dimensional Dirac fermions in $SrMnBi_2$. *Phys. Rev. B* **84,** 220401(R) (2011).
12. Wang, K. *et al.* Two-dimensional Dirac fermions and quantum magnetoresistance in $CaMnBi_2$. *Phys. Rev. B* **85,** 041101(R) (2012).
13. He, J. B., Wang, D. M. & Chen, G. F. Giant magnetoresistance in layered manganese pnictide $CaMnBi_2$. *Appl. Phys. Lett.* **100,** 112405 (2012).
14. Wang, J. *et al.* Layered transition-metal pnictide $SrMnBi_2$ with metallic blocking layer. *Phys. Rev. B* **84,** 064428 (2011).
15. Lee, G., Farhan, M. A., Kim, J. S. & Shim, J. H. Anisotropic Dirac electronic structures of $AMnBi_2$ (A=Sr, Ca). *Phys. Rev. B* **87,** 245104 (2013).
16. Feng, Y. *et al.* Strong anisotropy of Dirac cones in $SrMnBi_2$ and $CaMnBi_2$ revealed by angle-resolved photoemission spectroscopy. *Sci. Rep.* **4,** 5385 (2014).
17. Jia, L. L. *et al.* Observation of well-defined quasiparticles at a wide energy range in a quasi-two-dimensional system. *Phys. Rev. B* **90,** 035133 (2013).
18. Guo, Y. F. *et al.* Coupling of magnetic order to planar Bi electrons in the anisotropic Dirac metals

AMnBi$_2$ (A = Sr, Ca). *Phys. Rev. B* **90,** 075120 (2014).

19. Calder, S. *et al.* Magnetic structure and spin excitations in BaMn$_2$Bi$_2$. *Phys. Rev. B* **89,** 064417 (2014).

20. Cottam, M. G. & Lockwood, D. J. *Light Scattering in Magnetic Solids* Ch. 6 (John Wiley & Sons., New York, 1986).

21. Cardona, M. & Güntherodt, G. *Light Scattering in Solids IV* Ch. 4 (Springer-Verlag, Berlin, 1984).

22. Saparov, B. & Sefat, A. S. Crystals, magnetic and electronic properties of a new ThCr$_2$Si$_2$-type BaMn$_2$Bi$_2$ and K-doped compositions. *J. Solid State Chem.* **204,** 32-39 (2013).

23. Legrand, E. & Plumier, R. Neutron diffraction investigation of antiferromagnetic K$_2$NiF$_4$. *Phys. Status Solidi B* **2,** 317-320 (1962).

24. de Wijn, H. W., Walker, L. R. & Walstedt, R. E. Spin-wave analysis of the quadratic-layer antiferromagnets KNiF$_4$, KMnF$_4$ and RbMnF$_4$. *Phys. Rev. B* **8,** 285-295 (1973).

25. Devereaux, T. P. & Hackl, R. Inelastic Light Scattering from Correlated Electrons. *Rev. Mod. Phys.* **79**, 175 (2007).

## Acknowledgements

This work was supported by the Ministry of Science and Technology of China (Grant Nos.: 2016YFA0300504, 2016YFA0300501 and 2016YFA0300604) and the NSF of China. C. F. C. was supported in part by DOE under Cooperative Agreement No. DENA0001982. Y.G.S was supported by the Strategic Priority Research Program (B) of the Chinese Academy of Sciences (Grant No. XDB07020100). Q.M.Z., A.M.Z. and T.L.X were supported by the Fundamental Research Funds for the Central Universities and the Research Funds of Renmin University of China (10XNI038, 14XNLF06, 14XNLQ07).

## Author contributions

Q.M.Z conducted the whole study and wrote the paper. A.M.Z made Raman measurements, data analysis and wrote the paper. C.L.L made numerical calculations and data analysis. R.Y made theoretical modeling and calculations, and wrote the paper. C.J.Y and Y.G.S grew SrMnBi$_2$ and CaMnBi$_2$ single crystals. G.H.Z and T.L.X grew BaMn$_2$Bi$_2$ single crystals. C.F.C, X.Q.W. and J.T.J made data analysis and paper revising and checking.

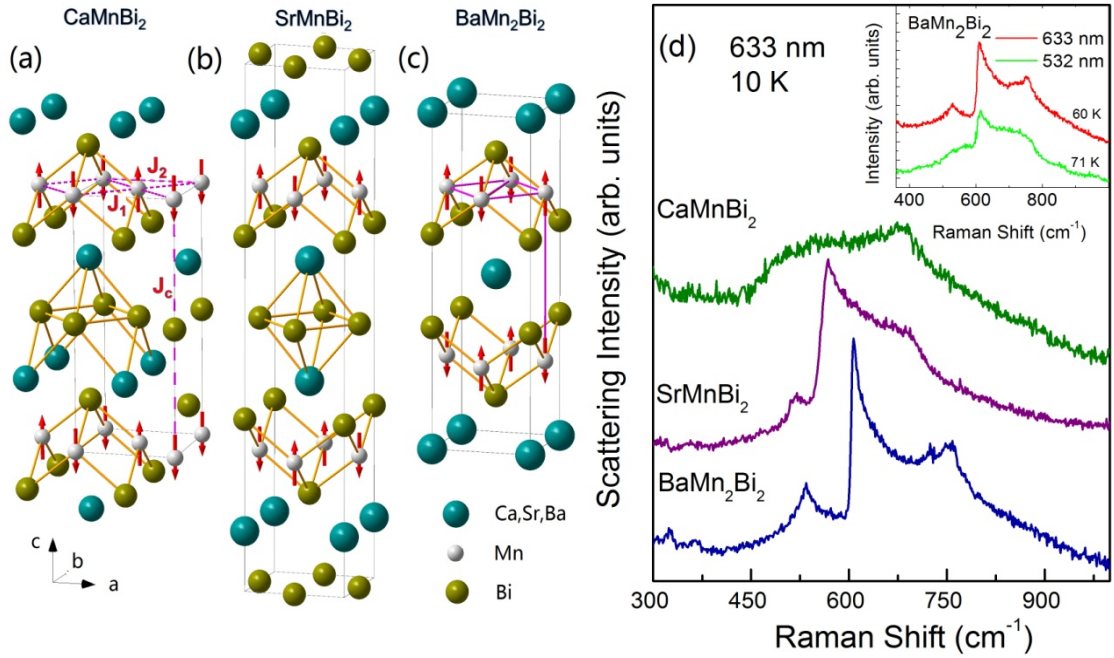

Figure 1 **Crystal and magnetic structures and two-magnon Raman spectra of CaMnBi₂, SrMnBi₂ and BaMn₂Bi₂.** (a)-(c) The crystal and magnetic structures of the three compounds[18,19]. (d) Their two-magnon excitations measured at 10 K. The inset shows the two-magnon spectra at different excitation energies. The spectra were collected in an unpolarized configuration to obtain a better signal-to-noise ratio.

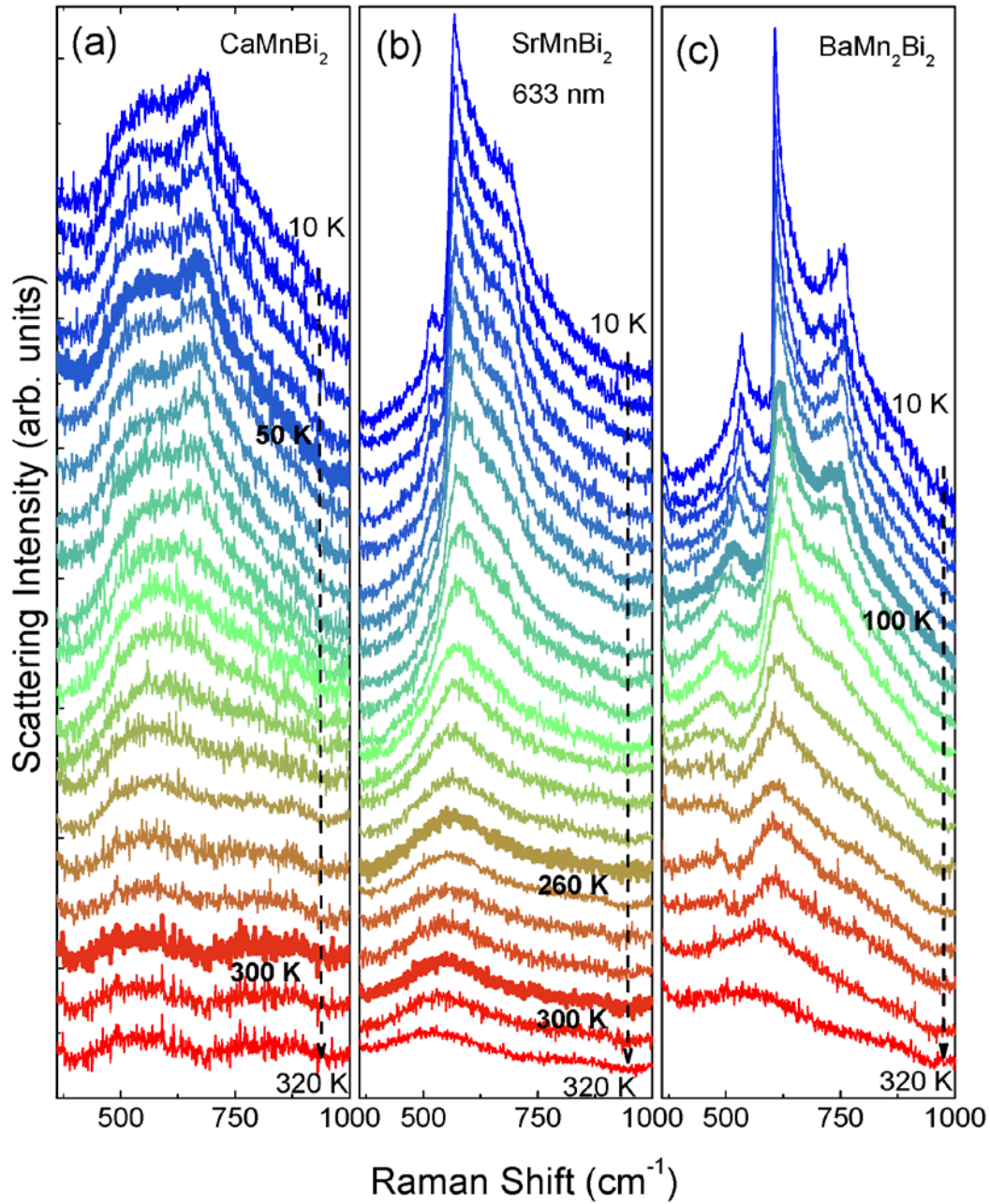

Figure 2 **Temperature evolution of the two-magnon spectra.** Measured Raman spectra of (a) CaMnBi$_2$, (b) SrMnBi$_2$, and (c) BaMn$_2$Bi$_2$. Highlighted are the spectra at 50 K and 300 K in (a), 260 K and 300 K in (b) and at 100 K in (c) where resistivity and susceptibility anomalies or antiferromagnetic transitions were observed in measurements using other techniques as discussed in the text, but no anomalies are visible in the two-magnon spectra at these temperatures. The spectra were collected in an unpolarized configuration to obtain a better signal-to-noise ratio.

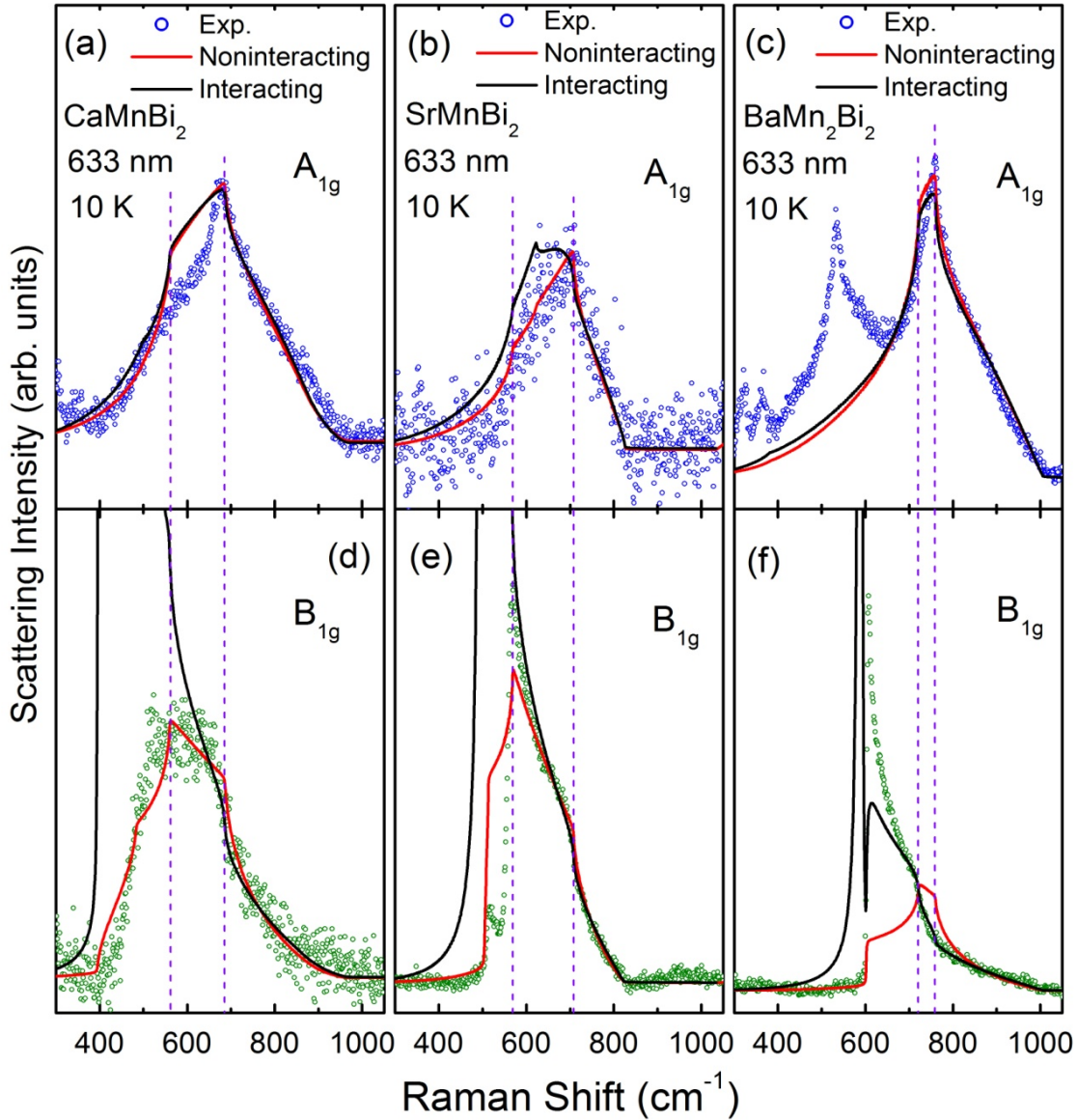

Figure 3 **Measured and calculated two-magnon Raman spectra.** The low-temperature (10 K) two-magnon Raman spectra in the $A_{1g}$ and $B_{1g}$ channels (circles) compared to the calculated results based on the linear spin-wave theory without considering the magnon-magnon interactions (red lines) and the results taking into account the magnon-magnon interactions (black lines). The vertical dashed lines mark the characteristic frequencies that are independent of the symmetries and the magnon-magnon interactions. These two characteristic frequencies, in combination with the cut-off frequencies, determine the exchange energies $J_1$, $J_2$ and $J_c$ (see Supplementary Note 2 and Supplementary Fig. 1 for details).

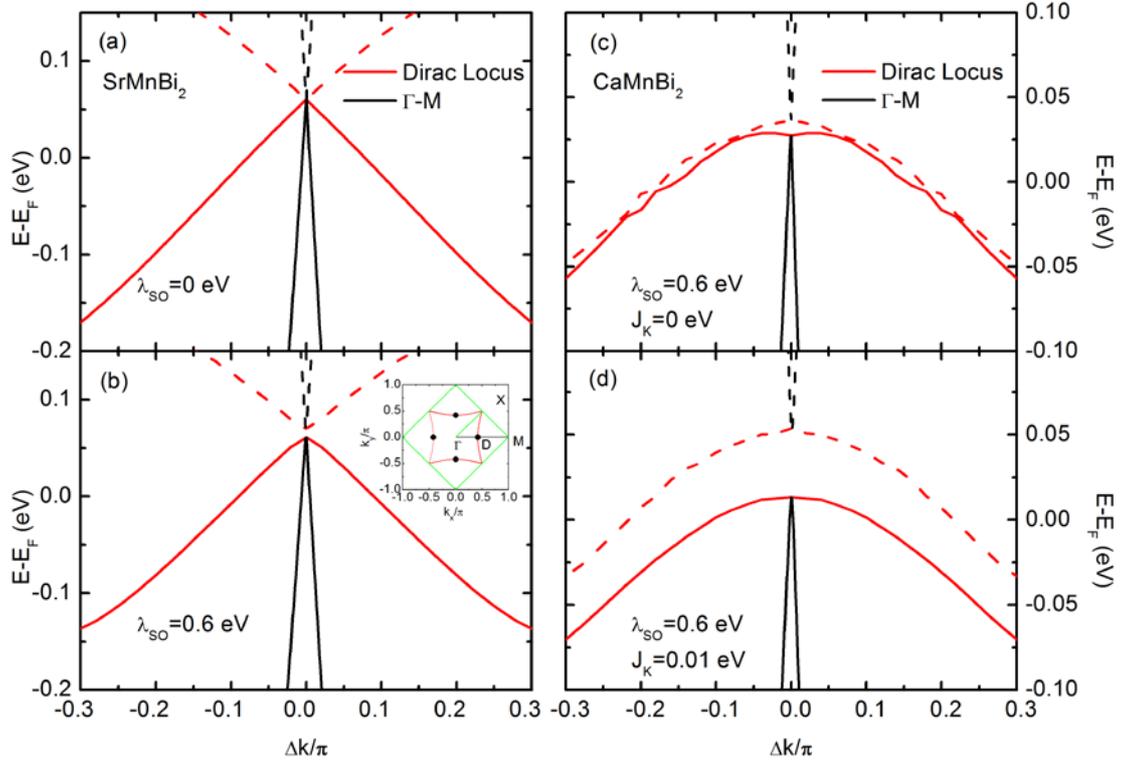

Figure 4 **Anisotropic Dirac bands in SrMnBi$_2$ and CaMnBi$_2$.** Panels (a) and (b) show the dispersion of the Dirac bands (black lines) along the Γ-M direction of the Brilluion zone and the locus of the crossing point energy of the Dirac bands (red lines) without or with the spin-orbit coupling $\lambda_{SO}$ for SrMnBi$_2$. The lower and upper branches are shown in solid and dashed lines. The inset in panel (b) shows the isolated Dirac points (black dots D) in SrMnBi$_2$ and continuous Dirac points (Dirac locus, red line) in CaMnBi$_2$. Panels (c) and (d) show the corresponding Dirac bands (black lines) along the Γ-M direction and the Dirac locus (red lines) with the spin-orbit coupling $\lambda_{SO}$ for CaMnBi$_2$ at two different values of the exchange coupling $J_K$. Here, the gap size between the lower and upper Dirac bands are dominated by the exchange coupling $J_K$ instead of the spin-orbit coupling $\lambda_{SO}$. Here $\Delta k=|k-k_D|$, where $k_D$ is the moment of the Dirac point. The Γ-M direction and Dirac locus direction are along the black line and red curve in the inset in panel (b), respectively.

Supplementary Information

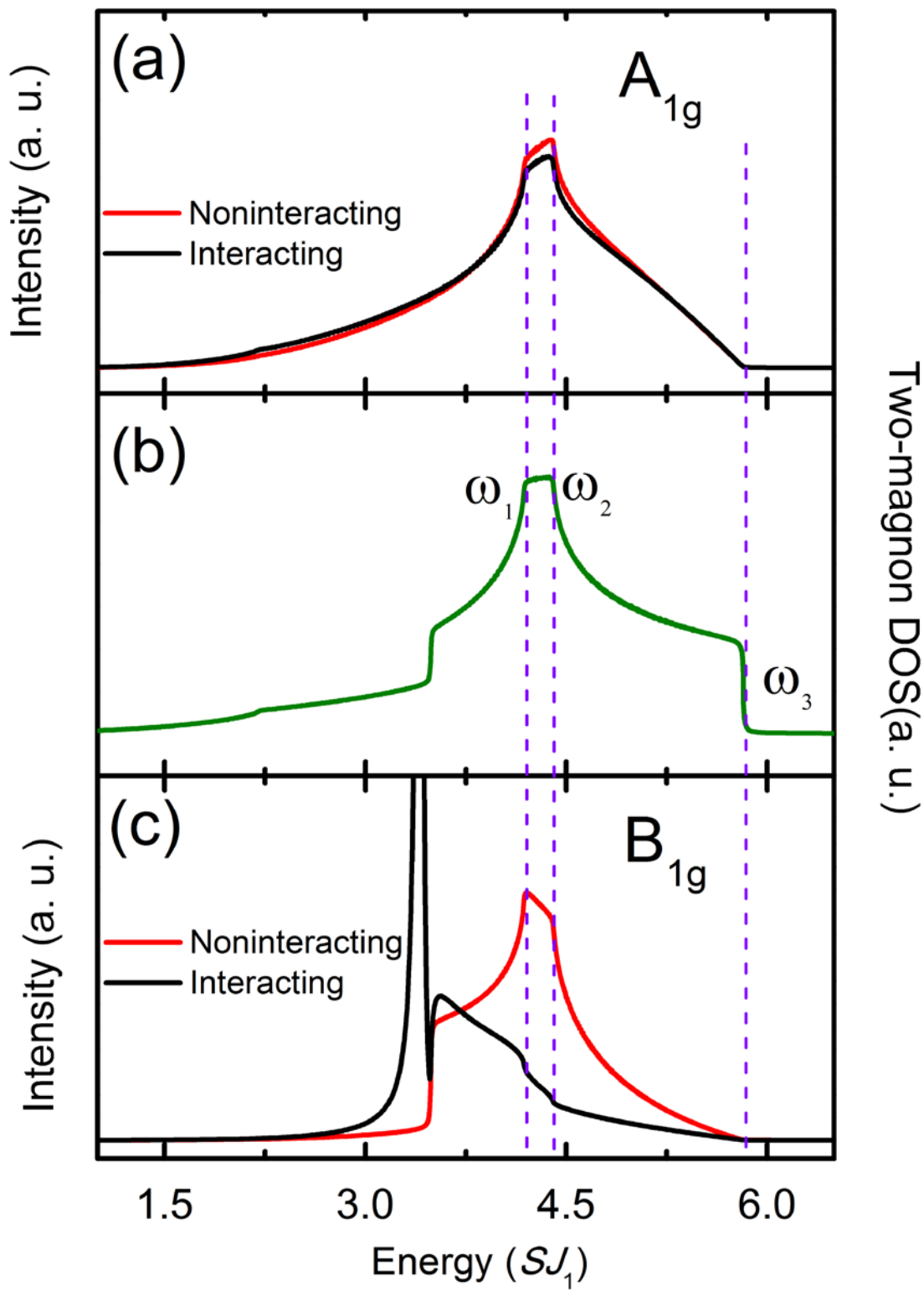

**Supplementary Figure 1**: The three characteristic frequencies $\omega_1$, $\omega_2$ and $\omega_3$ in the two-magnon Raman spectra and van-Hove singularities in the two-magnon DOS.

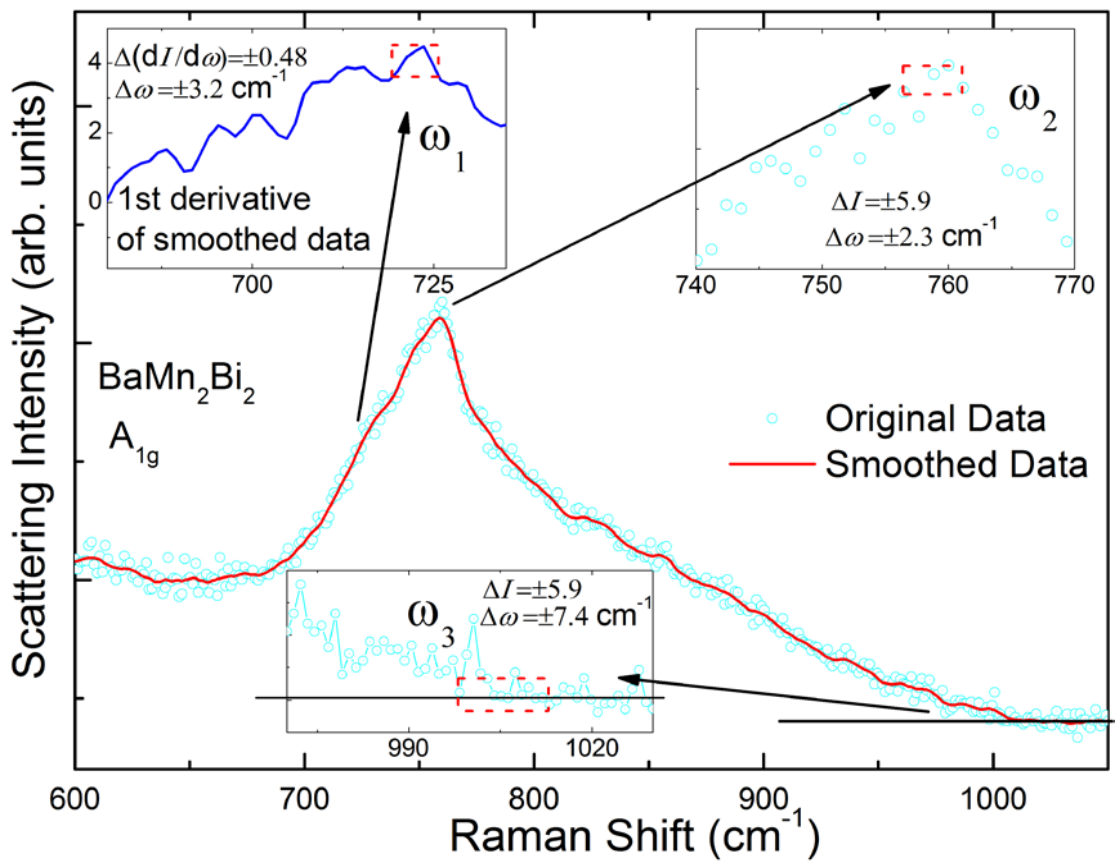
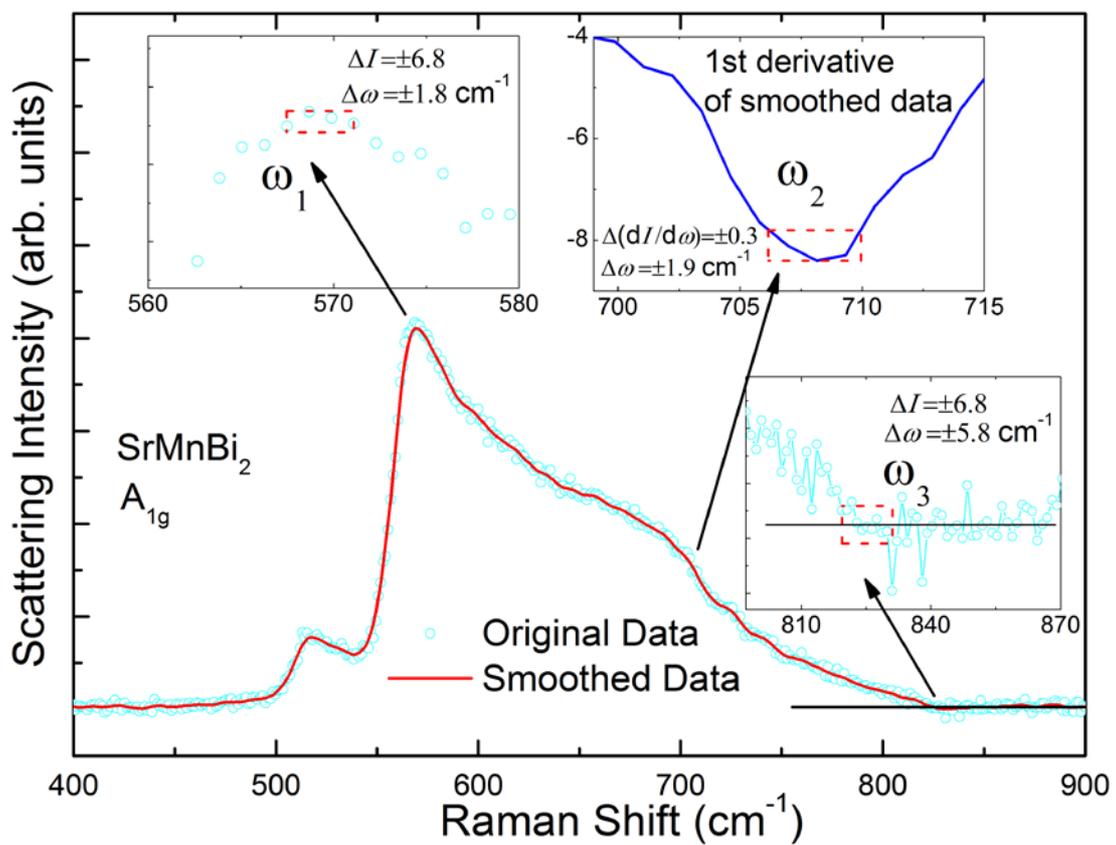

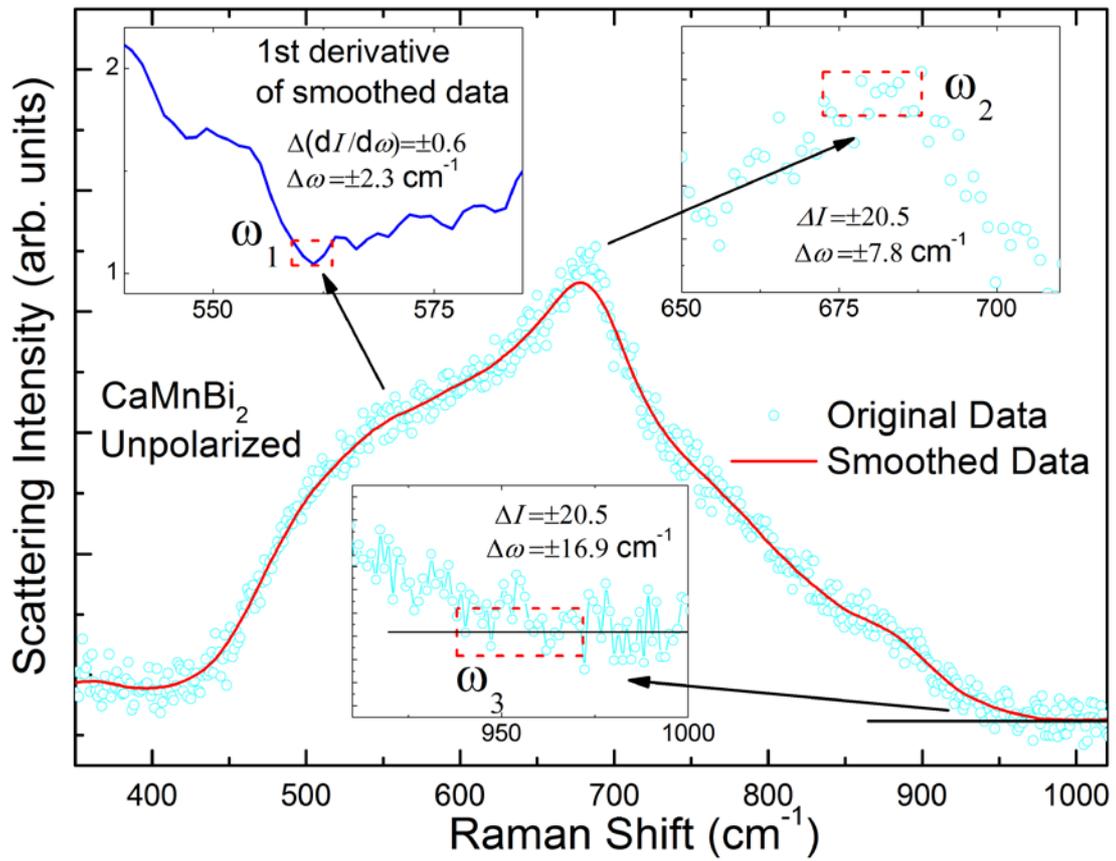

**Supplementary Figure 2: Determination of characteristic spectral points in BaMn$_2$Bi$_2$ (upper), SrMnBi$_2$ (medium) and CaMnBi$_2$ (lower).** The red dashed boxes in all the insets indicate the error ranges, as described in the above text. The black lines are the baselines given by fitting the background at the high-frequency end.

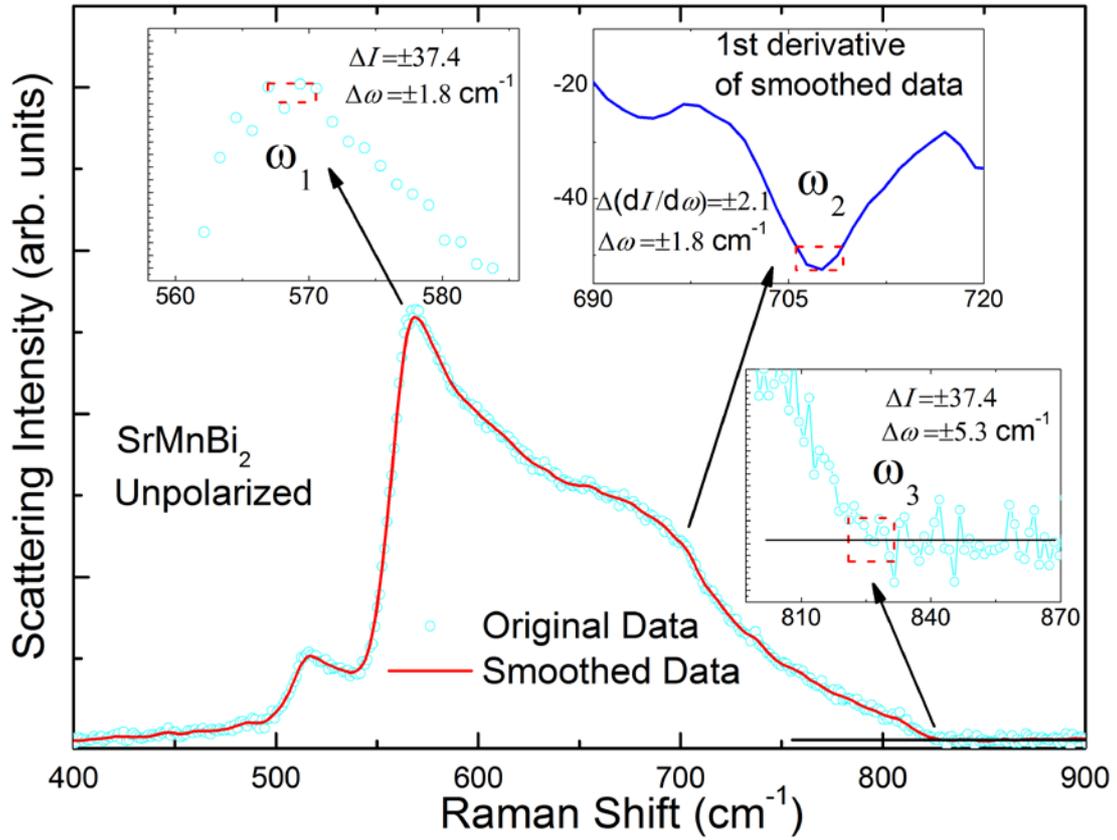

**Supplementary Figure 3: Determination of characteristic spectral points in SrMnBi$_2$ using the unpolarized spectra**. The red dashed boxes in the insets indicate the error ranges.

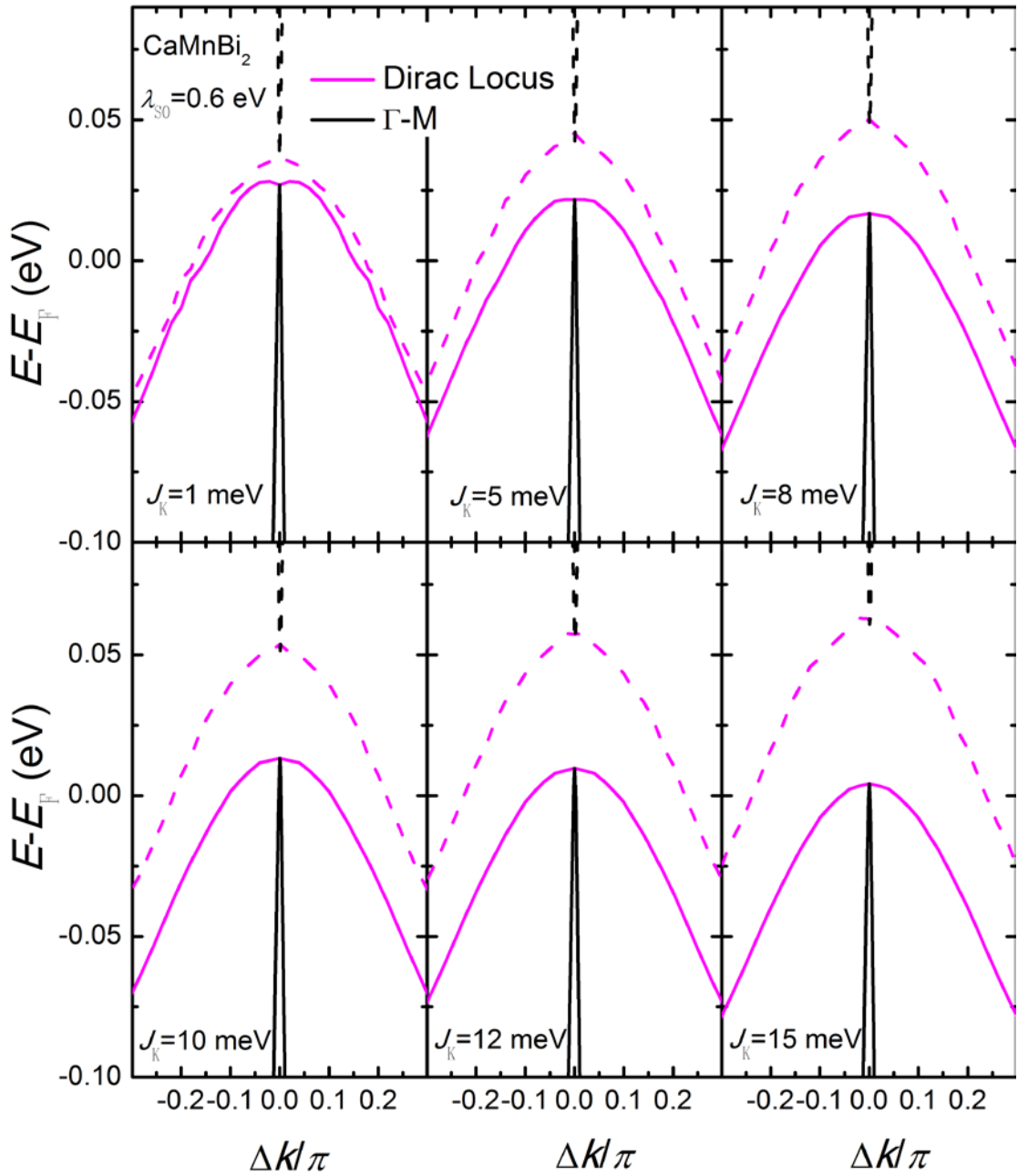

**Supplementary Figure 4: The energy gaps of anisotropic Dirac bands with various $J_k$ in CaMnBi$_2$.** The figure shows the dispersion of the Dirac bands (black lines) along the Γ-M direction of the Brilluion zone and the locus of the crossing point energy of the Dirac bands (magenta lines) for CaMnBi$_2$. The lower and upper branches are shown in solid and dashed lines.

**Supplementary Table 1: Three characteristic frequencies obtained from the experimental spectra and the extracted exchange parameters with the error bars obtained through the error estimation described in Supplementary Note 3.**

|  | CaMnBi$_2$ | SrMnBi$_2$ | BaMn$_2$Bi$_2$ |
|---|---|---|---|
| $\omega_1$ (cm$^{-1}$) | 561.2(2.3) | 569.3(1.8) | 722.6(3.2) |
| $\omega_2$ (cm$^{-1}$) | 680.2(7.8) | 708.1(1.9) | 758.7(2.3) |
| $\omega_3$ (cm$^{-1}$) | 954.9(16.9) | 825.2(5.8) | 1005.5(7.4) |
| $SJ_1$ (meV) | 20.77(0.79) | 16.00(0.30) | 21.45(0.32) |
| $SJ_2$ (meV) | 7.29(0.48) | 4.75(0.17) | 6.26(0.20) |
| $SJ_c$ (meV) | -1.31(0.10) | 2.92(0.09) | 0.78(0.08) |

**Supplementary Table 2: Comparison of the parameters in SrMnBi$_2$ extracted from the polarized and unpolarized spectra, respectively.**

| SrMnBi$_2$ | Polarized (A$_{1g}$) | Unpolarized |
|---|---|---|
| $\omega_1$ (cm$^{-1}$) | 569.2(1.8) | 568.7(1.8) |
| $\omega_2$ (cm$^{-1}$) | 708.1(1.9) | 707.3(1.8) |
| $\omega_3$ (cm$^{-1}$) | 825.2(5.8) | 826.2(5.3) |
| $SJ_1$ (meV) | 16.00(0.30) | 16.08(0.27) |
| $SJ_2$ (meV) | 4.75(0.17) | 4.79(0.15) |
| $SJ_c$ (meV) | 2.92(0.09) | 2.89(0.08) |

**Supplementary Table 3: Comparison of exchange interactions for three materials with/without single-ion anisotropy.**

|  |  | CaMnBi$_2$ | SrMnBi$_2$ | BaMn$_2$Bi$_2$ |
|---|---|---|---|---|
| $\frac{D}{J_1} = 0.046$ | $SJ_1$ (meV) | 20.85(0.81) | 15.97(0.39) | 21.24(0.32) |
|  | $SJ_2$ (meV) | 7.76(0.51) | 4.97(0.23) | 6.51(0.21) |
|  | $SJ_c$ (meV) | -1.23(0.10) | 2.70(0.10) | 0.72(0.07) |
| $\frac{D}{J_1} = 0$ | $SJ_1$ (meV) | 20.77(0.79) | 16.00(0.30) | 21.45(0.32) |
|  | $SJ_2$ (meV) | 7.29(0.48) | 4.75(0.17) | 6.26(0.20) |
|  | $SJ_c$ (meV) | -1.31(0.10) | 2.92(0.09) | 0.78(0.08) |

**Supplementary Table 4: The hopping parameters and the chemical potential $\mu$**

| Model parameters | SrMnBi$_2$ (eV) | CaMnBi$_2$ (eV) |
|---|---|---|
| $t_{1\sigma}^x$ | 2.00 | 2.00 |
| $t_{1\pi}^x$ | -0.50 | -0.50 |
| $t_2^x$ | 0.12 | 0.26 |
| $\mu$ | 0.00 | 0.11 |

## Supplementary Note 1: Method of calculation for two-magnon Raman scattering spectra

The standard theory of magnetic Raman scattering is based on the Elliott-Fleury-London theory[1]. The Raman scattering operator is given by

$$\hat{O} \sim \sum_{ij} J_{ij} (\hat{\mathbf{e}}_{in} \cdot \hat{\mathbf{d}}_{ij})(\hat{\mathbf{e}}_{out} \cdot \hat{\mathbf{d}}_{ij}) \, \mathbf{S}_i \cdot \mathbf{S}_j. \qquad (1)$$

Here $J_{ij}$ is the spin exchange interaction between the local moments at site $i$ and site $j$, $\hat{\mathbf{e}}_{in}$ and $\hat{\mathbf{e}}_{out}$ are unit polarization vectors of the incoming and scattering light, respectively, and $\hat{\mathbf{d}}_{ij}$ is the vector connecting site $i$ and site $j$. The Raman cross section at zero temperature is given by the imaginary part of the correlation function $I(\omega) = -i \int \mathrm{d}t \, e^{i\omega t} \langle \mathcal{T}_t \hat{O}^+(t) \hat{O}(0) \rangle_0$, where $\langle ... \rangle_0$ represents the quantum mechanical average over the ground state, and $\mathcal{T}_t$ is the time ordering operator.

We use the spin-wave approach within the framework of perturbative theory to calculate Raman spectra[2]. By introducing the Holstein-Primakoff transformation, we express spins in A (spin up) and B (spin down) sublattices in terms of H-P boson operators $a_i$ and $b_j$. The system Hamiltonian $\mathrm{H} = J_1 \sum_{\langle ij \rangle} \mathbf{S}_i \cdot \mathbf{S}_j + J_2 \sum_{\langle\langle ij \rangle\rangle} \mathbf{S}_i \cdot \mathbf{S}_j + J_c \sum_{\langle\langle\langle ij \rangle\rangle\rangle} \mathbf{S}_i \cdot \mathbf{S}_j$ is expanded in powers of $1/S$ as:

$$\mathrm{H} \approx S^2 \left( E_0 + \frac{1}{S} \hat{H}_0 + \frac{1}{S^2} \hat{H}_1 + \cdots \right) \qquad (2)$$

Here $E_0$ is a constant classical energy, $\hat{H}_0$ contains quadratic terms of magnons and represents the linear spin-wave (LSW) correction to the classical energy, and $\hat{H}_1$ contains magnon quartic terms, representing two-body magnon-magnon interactions. The higher order terms are ignored in the present treatment.

To calculate the correlation function $I(\omega)$, we first apply Fourier and Bogoliubov transformations to diagonalize the quadratic LSW part $\hat{H}_0$ in terms of Bogoliubov magnons $\alpha_{\mathbf{k}}$ and $\beta_{-\mathbf{k}}$. The $\hat{H}_1$ part then contains Bogoliubov magnons in quartic order, and the magnon-pair-scattering term $\alpha_{\mathbf{k}}^+ \beta_{-\mathbf{k}}^+ \beta_{-\mathbf{k}'} \alpha_{\mathbf{k}'}$ in $\hat{H}_1$ is treated within the ladder approximation. The interaction vertex is reserved to the lowest $(1/S)^0$ order, and the ladder diagrams are summed up exactly.

## Supplementary Note 2: Determination of exchange energies $J_1$, $J_2$ and $J_c$

We determine the exchange couplings $J_1$, $J_2$ and $J_c$ from three characteristic frequencies $\omega_1$, $\omega_2$ and $\omega_3$, which, as shown in Supplementary Figure 1, correspond to the frequencies of the shoulders and cut-off of the $A_{1g}$ Raman spectra. These frequencies are associated with the van-Hove singularities in the two-magnon density of states (DOS), and are not shifted by the magnon-magnon interactions within our approximations (Supplementary Figure 1). In the linear spin-wave theory, these frequencies take the following analytical form:

1) G-type AFM

$$\omega_1 = 4S(2J_1 + J_c)\sqrt{\frac{J_1-2J_2}{J_1+2J_2}} \quad (3)$$

$$\omega_2 = \frac{4S[2J_1(J_1-2J_2)+J_c(J_1+2J_2)]}{\sqrt{J_1^2-4J_2^2}} \quad (4)$$

$$\omega_3 \gtrsim 4S\sqrt{\frac{(2J_1-2J_2+J_c)[2(J_1-J_2)(J_1^2-2J_1J_2+2J_2^2)+J_c(J_1-2J_2)^2]}{J_1^2-2J_1J_2+J_2(2J_2-J_c)}} \quad (5)$$

$$\omega_2 - \omega_1 = SJ_c \times \frac{16J_2/J_1}{\sqrt{1-(2J_2/J_1)^2}} \quad (6)$$

which holds when $J_c < 4J_2$ and $J_c < \frac{2J_1(J_1-2J_2)}{J_1+2J_2}$;

2) C-type AFM

$$\omega_1 = 8SJ_1\sqrt{\frac{J_1-2J_2}{J_1+2J_2}} \quad (7)$$

$$\omega_2 = \frac{8SJ_1(J_1-2J_2-J_c)}{\sqrt{J_1^2-4J_2^2}} \quad (8)$$

$$\omega_3 = 8S(J_1 - J_2 - J_c) \quad (9)$$

$$\omega_2 - \omega_1 = |SJ_c| \times \frac{8}{\sqrt{1-(2J_2/J_1)^2}} \quad (10)$$

which holds when $|J_c| < \frac{J_1(J_1-2J_2)}{2J_2}$, and these constraints for both cases are valid in the cases studied in the present work. From the experimentally determined frequencies, we can extract the three exchange couplings $J_1$, $J_2$ and $J_c$ using the above formulas.

**Supplementary Note 3: Error estimation of the exchange energies**

The characteristic frequencies can be expressed as
$$\boldsymbol{\omega} = \boldsymbol{\omega}(\mathbf{J}) \quad (11)$$
The experimentally determined frequencies have errors around their averages
$$\boldsymbol{\omega} = \bar{\boldsymbol{\omega}} + \delta\boldsymbol{\omega} \quad (12)$$
The corresponding exchange energies can then be written as $\mathbf{J} = \bar{\mathbf{J}} + \delta\mathbf{J}$, where $\bar{\mathbf{J}}$ is determined via the equation $\bar{\boldsymbol{\omega}} = \boldsymbol{\omega}(\bar{\mathbf{J}})$, and $\delta\mathbf{J}$ is obtained through the expansion

$$\delta\boldsymbol{\omega} = \left(\frac{\partial\boldsymbol{\omega}}{\partial\mathbf{J}}\right)\Big|_{\mathbf{J}=\bar{\mathbf{J}}} \delta\mathbf{J}, \quad (13)$$

where $\left(\frac{\partial\boldsymbol{\omega}}{\partial\mathbf{J}}\right)\Big|_{\mathbf{J}=\bar{\mathbf{J}}}$ is the Jacobi's determinant, which leads to

$$\delta\mathbf{J} = \left[\left(\frac{\partial\boldsymbol{\omega}}{\partial\mathbf{J}}\right)\Big|_{\mathbf{J}=\bar{\mathbf{J}}}\right]^{-1} \delta\boldsymbol{\omega}. \quad (14)$$

**Supplementary Note 4: Effects of possible spin anisotropy on the exchange energies in SrMnBi$_2$ and CaMnBi$_2$**

In Sr(Ca)MnBi$_2$ materials Dirac carriers in Ca(Sr)Bi layers are subjected to SOC effect which may introduce spin anisotropy in the exchange couplings between Mn ions. To take into account the spin anisotropy effect we consider the following model

$$\text{H} = J_1 \sum_{\langle ij \rangle} [S_i^z S_j^z + \frac{\eta}{2}(S_i^+ S_j^- + S_i^- S_j^+)] + J_2 \sum_{\langle\langle ij \rangle\rangle} \mathbf{S}_i \cdot \mathbf{S}_j + J_c \sum_{\langle\langle\langle ij \rangle\rangle\rangle} \mathbf{S}_i \cdot \mathbf{S}_j, \quad (15)$$

where $\eta$ is the parameter describing the anisotropy, and $\eta < 1$ ensures that spins are aligned along the z direction.

To determine the parameters $J_1$, $J_2$, $J_c$ and $\eta$, we make use of an additional characteristic point in the measured Raman spectra, namely the absorption edge in the $B_{1g}$ channel. Following the same procedure described above, we find that compared to the isotropic model the difference in the exchanges energies caused by the anisotropy is less than 10%, which indicates that the results obtained using the isotropic model are quite reasonable.

**Supplementary Note 5: Effects of single-ion anisotropy in CaMnBi$_2$, SrMnBi$_2$ and BaMn$_2$Bi$_2$**

It has been reported in inelastic neutron scattering experiments that in BaMn$_2$Bi$_2$ materials there is a spin gap of 16 meV[4]. The magnetic excitations can be well fit to a Heisenberg model plus single-ion anisotropy terms[4]

$$\text{H} = \sum_{ij} J_{ij} \mathbf{S}_i \cdot \mathbf{S}_j - D \sum_i (S_i^z)^2, \quad (31)$$

where $D > 0$ ensures that spins are aligned along the *z* direction.
We find that all characteristic frequencies will be shifted in presence of finite $D$ terms. However, the relation

$$\omega_2 - \omega_1 = SJ_c \times \frac{16 J_2/J_1}{\sqrt{1-(2J_2/J_1)^2}} \quad (32)$$

for G-type AFM and

$$\omega_2 - \omega_1 = |SJ_c| \times \frac{8}{\sqrt{1-(2J_2/J_1)^2}} \quad (33)$$

for C-type AFM still remains unchanged. This implies that the width of the shoulders in $A_{1g}$ spectra still has no direct relations with $D$ terms. Using the value $D/J_1 = 0.046$ extracted in INS experiments for BaMn$_2$Bi$_2$ materials[4], we can calculate the exchange interactions as is shown in Supplementary Table 3.

We can see that in presence of $D$ terms, the exchange interactions are slightly changed. The interchange coupling $SJ_c$ becomes slightly smaller in presence of $D$, but the difference is less than 10%. It is not surprising that $D$ terms can generate a relatively big spin gap but do not strongly affect the exchange interaction results extracted from Raman experiments, since $D$ terms mainly affect low energy magnetic excitations, but Raman techniques mainly probe high energy physics.

**Supplementary Note 6: Determination of characteristic frequencies and their errors**

The spectra for $BaMn_2Bi_2$ and $SrMnBi_2$ have a good signal-to-noise level, and the frequencies can be directly read out from the raw spectra ($\omega_2$ in $BaMn_2Bi_2$ and $\omega_1$ in $SrMnBi_2$) or obtained from taking the first derivative ($\omega_1$ in $BaMn_2Bi_2$ and $\omega_2$ in $SrMnBi_2$). For $CaMnBi_2$, one must be more careful as the original polarized spectra have a higher noise level, making it hard to accurately determine the characteristic spectral points. On the other hand, spin-wave calculations indicate that both polarized spectra ($A_{1g}$ and $B_{1g}$) have exactly the same characteristic points. This means that the combined $A_{1g}$ and $B_{1g}$ spectra, i.e., the unpolarized data, possess the same characteristic spectral points. This enables an alternative scenario, where we have collected many unpolarized spectra at 10 K with a much better signal-to-noise ratio, as shown in Supplementary Figure 2. Using the unpolarized spectra ($A_{1g}+B_{1g}$), one can easily determine the characteristic frequencies and their errors, in the same way as done above for $BaMn_2Bi_2$ and $SrMnBi_2$. We have examined the validity of the method using $SrMnBi_2$, in which there is little difference between the parameters derived from the polarized spectra and the unpolarized ones, respectively (Supplementary Note 7).

Then we can extract the frequency parameters and quantitatively estimate the associated errors in the following procedure (see Supplementary Figure 2). 1) A general linear fitting was made for a linear region selected from the raw spectra or the first-derived data. This gives the standard deviations in intensity (or its derivative); 2) For $\omega_1$ and $\omega_2$ associated with the local maximums/minimums of the raw spectra or the first-derivatives, we can identify a region which starts from the maximums/minimums and vertically extends by double standard deviations (the heights of the red dashed error boxes). All the data points in the region are possible maximums/minimums. This simply fixes the corresponding deviations in frequency (the widths of the error boxes) and provides the standard errors; 3) For $\omega_3$ not associated with a local extreme, we first determined the baselines (black lines) by fitting the background at the high-frequency end. The left bound of the error box is reached when the positive intensity deviations from a baseline begin to exceed the standard deviations estimated in the first step. Similarly the right bound is defined at the position where the intensity deviations approach the standard ones. This gives the frequency parameter $\omega_3$ and the associated errors. The characteristic frequencies obtained from the experimental spectra strictly following the above procedure and the extracted exchange parameters with the error bars, are listed in Supplementary Table 1.

**Supplementary Note 7: Comparison of the parameters extracted from the polarized and unpolarized spectra in SrMnBi$_2$**

Following the procedure described in Supplementary Note 6, we can also obtain the frequency parameters in SrMnBi$_2$ with the unpolarized spectra (see Supplementary Figure 3). And the obtained characteristic frequencies and the extracted exchange parameters from the polarized and unpolarized spectra, are listed in Supplementary Table 2 for comparison. There is little difference between both cases. This demonstrates that the unpolarized spectra work well as the polarized ones in obtaining the characteristic frequency points.

**Supplementary Note 8: Electronic band structures and Dirac points in SrMnBi$_2$ and CaMnBi$_2$**

We consider the spin-fermion Hamiltonian introduced in the main text

$$H = \sum_{i,j,\alpha,\beta,l} t_{ij}^{\alpha\beta l} c_{i\alpha l}^+ c_{j\beta l} + \lambda_{SO} \sum_{i\alpha\beta ll'} c_{i\alpha l}^+ c_{i\beta l'} + \frac{J_K}{2} \sum_{i\alpha ll'} c_{i\alpha l}^+ \sigma_{ll'} c_{j\alpha l'} \cdot \mathbf{S}_{i\pm\hat{z}} + \sum_{i'j'} J_{i'j'}^H \mathbf{S}_{i'} \cdot \mathbf{S}_{j'} \quad (16)$$

The first term is a two-orbital tight-binding model for the itinerant electrons in Bi 6$p_x$ and 6$p_y$ orbits (in the Sr(Ca)Bi layer). The second term contains the atomic spin-orbit coupling. $\mathbf{S}_{i\pm\hat{z}}$ refers to the local moment of Mn in layers above or below the Bi site i. For simplicity, we do not consider the influence of the Bi 6$p_z$ orbit. Since the observed magnetic moments in these materials are about 4 $\mu_B$ per Mn, we treat $\mathbf{S}_{i'}$ as classical spins. We take the $J_{i'j'}^H$ values obtained from the Raman measurements, so that the ground state of the model has either a G-AFM (SrMnBi$_2$) or a C-AFM (CaMnBi$_2$) order. We then treat the effects of the AFM order on the band structure of the itinerant electrons at the mean-field level.

Within above approximations, the Mn local moments serve as local magnetic fields that couple to the itinerant electrons and modify their dispersion. The Hamiltonian is then written as

$$H \approx H_{TB} + H_S, \quad (17)$$

where

$$H_{TB} = \sum_{ij\alpha\beta l} t_{ij}^{\alpha\beta l} c_{i\alpha l}^+ c_{j\beta l} + \lambda_{SO} \sum_{i\alpha\beta ll'} c_{i\alpha l}^+ c_{i\beta l'}, \quad (18)$$

$$H_S = \frac{J_K}{2} \sum_{n\alpha} m_n (c^+_{n\alpha\uparrow} c_{n\alpha\uparrow} - c^+_{n\alpha\downarrow} c_{n\alpha\downarrow}), \qquad (19)$$

and $m_n = \langle S_n^z \rangle$ is the sublattice magnetic moment for n = A, B sublattice. For SrMnBi$_2$, the magnetic order is G-AFM, where the Mn ions in upper and lower layers belong to the A and B sublattices, respectively, and at the mean-field level, the effect from the two layers cancels out exactly, hence H ≈$H_{TB}$. But for CaMnBi$_2$, the C-AFM order induces an uncompensated magnetic field, which acts as a mass term since it has different signs on the two sublattices.

We rewrite H$_{TB}$ into a matrix form[3]:

$$H_{TB} = \begin{pmatrix} H_{AA} + H_{so} & H_{AB} \\ H_{BA} & H_{BB} + H_{so} \end{pmatrix}, \qquad (20)$$

where $H_{so}$ and $H_{nn'}$ are 2×2 matrices. The spin-orbit coupling takes the form

$$H_{so} = \begin{pmatrix} 0 & -i\lambda_{so} \\ i\lambda_{so} & 0 \end{pmatrix}, \qquad (21)$$

where $\lambda_{so}$ is the coupling constant. In our calculations, we consider two cases, $\lambda_{so} = 0$, and $\lambda_{so} = 0.6$ eV.[3]

For the hopping integrals, we assume that the dominant terms are the intraorbital ones, and we neglect the interorbital hopping. We then obtain the following hopping matrices for SrMnBi$_2$ and CaMnBi$_2$:

$$H^{Sr}_{AA} = \begin{pmatrix} 2t^{xs}_2 \cos(k_x + k_y) - \mu & 0 \\ 0 & 2t^{xs}_2 \cos(k_x + k_y) - \mu \end{pmatrix}, \qquad (22)$$

$$H^{Sr}_{BB} = \begin{pmatrix} 2t^{xs}_2 \cos(k_x - k_y) - \mu & 0 \\ 0 & 2t^{xs}_2 \cos(k_x - k_y) - \mu \end{pmatrix}, \qquad (23)$$

$$H^{Sr}_{AB} = H^{Sr}_{BA} = \begin{pmatrix} 2(t^{xs}_{1\sigma} \cos k_x + t^{xs}_{1\pi} \cos k_y) & 0 \\ 0 & 2(t^{xs}_{1\sigma} \cos k_y + t^{xs}_{1\pi} \cos k_x) \end{pmatrix}, \qquad (24)$$

$$H^{Ca}_{AA} = H^{Ca}_{BB} = \begin{pmatrix} 4t^{xc}_2 \cos k_x \cos k_y - \mu & 0 \\ 0 & 4t^{xc}_2 \cos k_x \cos k_y - \mu \end{pmatrix}, \qquad (25)$$

$$H^{Ca}_{AB} = H^{Ca}_{BA} = \begin{pmatrix} 2(t^{xc}_{1\sigma} \cos k_x + t^{xc}_{1\pi} \cos k_y) & 0 \\ 0 & 2(t^{xc}_{1\sigma} \cos k_y + t^{xc}_{1\pi} \cos k_x) \end{pmatrix}. \qquad (26)$$

Here the hopping parameters and the chemical potential $\mu$ are determined by fitting to the DFT band structure, and their values are summarized in Supplementary Table 4. Note that due to the buckling of the Sr cations, $H^{Sr}_{AA} \neq H^{Sr}_{BB}$ for general **k**. The band structure can be obtained by diagonalizing the mean-field Hamiltonian. Without the spin-orbit coupling, for SrMnBi$_2$, the energy in each band reads

$$E^{Sr}_{1(2),\pm} =$$

$$2t^{xs}_2 \cos k_x \cos k_y \pm \sqrt{4t^{xs}_2 \sin^2 k_x \sin^2 k_y + 4\left(t^{xs}_{1\sigma(\pi)} \cos k_x + t^{xs}_{1\pi(\sigma)} \cos k_y\right)^2} \qquad (27)$$

By requiring

$$\sqrt{4t_2^{xs} \sin^2 k_x \sin^2 k_y + 4\left(t_{1\sigma(\pi)}^{xs} \cos k_x + t_{1\pi(\sigma)}^{xs} \cos k_y\right)^2} = 0, \text{ we obtain four Dirac}$$

points. For $t_{1\sigma}^{xs} > t_{1\pi}^{xs}$, they are located at $k_x = 0$, $k_y = \pm \arccos(-t_{1\pi}^{xs}/t_{1\sigma}^{xs})$, and $k_y = 0$, $k_x = \pm \arccos(-t_{1\pi}^{xs}/t_{1\sigma}^{xs})$. Focusing on one Dirac point $k_y = 0$, $k_x = \pm \arccos(-t_{1\pi}^{xs}/t_{1\sigma}^{xs})$, the dispersion is very anisotropic, as shown in Fig. 4(a) in the main text. Turning on the spin-orbit coupling pushes apart the Dirac points as shown in Fig. 4(b).

Following the same procedure, we obtain the electronic band structure for CaMnBi$_2$. Without the spin-orbit coupling,

$$E_{1(2),\pm}^{Ca} = 2t_2^{xc} \cos k_x \cos k_y \pm \sqrt{\tilde{J}_k^2 + 4\left(t_{1\sigma(\pi)}^{xc} \cos k_x + t_{1\pi(\sigma)}^{xc} \cos k_y\right)^2} \tag{28}$$

where $\tilde{J}_k = J_k |m_n|$ is the effective magnetic field to itinerant electrons due to the C-AFM order. If we neglect the magnetic coupling, the band structure has continuous Dirac points along the lines $k_x = \pm\arccos(-t_{1\pi}^{xc}/t_{1\sigma}^{xc} \cos k_y)$ and $k_y = \pm\arccos(-t_{1\pi}^{xc}/t_{1\sigma}^{xc} \cos k_x)$. The magnetic field term $\tilde{J}_k$ acts as a mass term and opens a gap between the upper and lower branches of the Dirac bands. Note that a finite spin-orbit coupling may also opens a gap, but it is much smaller compared to the one associated with the AFM order (see Fig.4(c)-(d), Supplementary Figure 4).

## Supplementary Note 9: Estimate of the RKKY interaction

In the above spin-fermion model, the coupling between itinerant electrons and local moments mediates an RKKY interaction among the local moments:

$$H_{RKKY}(\mathbf{R}_i - \mathbf{R}_j) = \frac{1}{2} J(\mathbf{R}_i - \mathbf{R}_j) \mathbf{S}_i \cdot \mathbf{S}_j, \tag{29}$$

with the RKKY coupling given by

$$J(\mathbf{R}) = 2\left(\frac{J_K}{N}\right)^2 \sum_{\mathbf{k},\mathbf{k}'} \frac{\Theta(E_\mathbf{k}) - \Theta(E_{\mathbf{k}'})}{E_\mathbf{k} - E_{\mathbf{k}'}} e^{i(\mathbf{k}-\mathbf{k}')\cdot \mathbf{R}}, \tag{30}$$

where $J(\mathbf{R}) = J(\mathbf{R}_i - \mathbf{R}_j)$, $\Theta(E_\mathbf{k})$ is the Fermi distribution function. A simple estimate based on the perturbation theory gives the induced RKKY coupling at the order of $(J_K)^2/E_F$, where $E_F$ is the Fermi energy of the relevant conduction band. Taking the value of $J_K$~10-20 meV, which is estimated from the size of the gap between the two Dirac bands, and $E_F$~0.1 eV from the fitting to the DFT results, we get that the RKKY-induced interlayer coupling $|J(\hat{z})|$~1-4 meV. These Jc values are compatible with those obtained from our Raman measurements (see Table I in the main text). This agreement with the experimental results suggests the model used in our analysis is valid in understanding the fundamental mechanisms underlying the novel phenomena observed in our measured Raman spectra. A more accurate quantitative description of the size and sign of the RKKY interaction would require more sophisticated band-structure calculations, which are beyond the scope of the present work.

## Supplementary References

1. Fleury, P. A. & Loudon, R. Scattering of light by one- and two-magnon excitations. *Phys. Rev.* **166,** 514-530 (1968).
2. Luo, C., Datta, T. & Yao, D. X. Spectrum splitting of bimagnon excitations in a spatially frustrated Heisenberg antiferromagnet revealed by resonant inelastic x-ray scattering. *Phys. Rev. B* **89,** 165103 (2014).
3. Lee, G., Farhan, M. A., Kim, J. S. & Shim, J. H. Anisotropic Dirac electronic structures of $AMnBi_2$ (A=Sr, Ca). *Phys. Rev. B* **87,** 245104 (2013).
4. Calder, S. *et al.* Magnetic structure and spin excitations in $BaMn_2Bi_2$. *Phys. Rev. B* **89,** 064417 (2014).